# Damage mechanisms in polyalkenes irradiated with ultra-short XUV/x-ray laser pulses

N. Nikishev[1,2], N. Medvedev[1,3,*]

1) Institute of Physics, Czech Academy of Sciences, Na Slovance 1999/2, 182 00 Prague 8, Czech Republic

2) Czech Technical University in Prague, Faculty of Nuclear Sciences and Physical Engineering, Břehová 7, 115 19 Prague 1, Czech Republic

3) Institute of Plasma Physics, Czech Academy of Sciences, Za Slovankou 3, 182 00 Prague 8, Czech Republic

## Abstract

Although polymers are widely used in laser-irradiation research, their microscopic response to high-intensity ultrafast XUV or X-ray irradiation is still largely unknown. Here we comparatively study homologous series of alkenes. XTANT-3 hybrid simulation toolkit is used to determine their damage kinetics and irradiation threshold doses. The code simultaneously models the nonequilibrium electron kinetics, the energy transfer between electrons and atoms *via* nonadiabatic electron-ion (electron-phonon) coupling, nonthermal modification of the interatomic potential due to electronic excitation, and the ensuing atomic response and damage formation. It is shown that the lowest damage threshold is associated with local defect creation such as dehydrogenation, various group detachment from the backbone, or polymer strand cross-linking. At higher doses, the disintegration of the molecules leads to a transient metallic liquid state: a nonequilibrium superionic state outside of the material phase diagram. We identify nonthermal effects as the leading mechanism of damage, whereas the thermal (nonadiabatic electron-ion coupling) channel influences the kinetics only slightly in the case of femtosecond-pulse irradiation. Despite notably different properties of the studied alkene polymers, the ultrafast-X-ray damage threshold doses are found to be very close to ~0.05 eV/atom in all three materials: polyethylene, polypropylene, and polybutylene.

---

[*] Corresponding author: nikita.medvedev@fzu.cz, Phone: +420 26605 2819

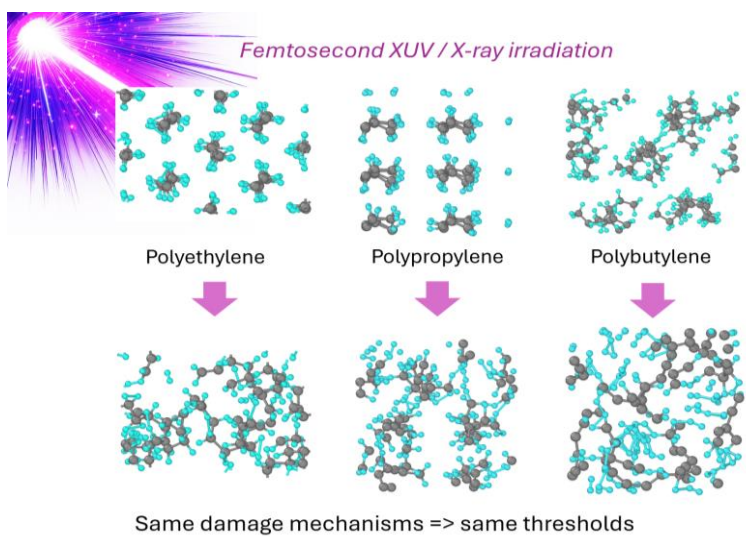

TOC Graphic

## I.    Introduction

Synthetic and natural polymers found vital and widespread applications in everyday activities [1,2]. Synthetic alkene polymers (polyolefins) display favourable characteristics for printability in complex architectures and show high biocompatibility of printable materials [3]. The response of alkene polymers to irradiation is used for practical applications in the processing of thin films in electronics [4], additive manufacturing [5], and sample analysis techniques [6].

Polyolefin-based materials show no degradation at doses below 10 kGy (~0.4 eV/atom) and may be used at cumulative doses up to 0.3 MGy[7], which makes them attractive candidates for practical use in conventional radiation scenarios (low dose rate) [8,9]. Under radiation exposure, polyethylene exhibits crosslinking, chain scission, and dehydrogenation, which lead to degradation [10–12]. However, the relative importance of the various damage processes may change depending on the dose rate [12]. The basic mechanisms of damage need to be investigated to understand the induced modifications of the polymer properties.

Free-electron lasers (FEL) produce intense femtosecond pulses of extreme ultraviolet (XUV)/X-ray radiation [13–15]. Due to the ultrashort pulse duration, such irradiation achieves extremely high dose rates. It enables the generation and examination of highly non-equilibrium states of matter in extreme conditions [16]. In particular, polymers are used at FEL facilities as pulse-shape monitors via methods of ablative importing [17–19]; as well as for modifying their properties and forming new polymers [20,21].

During the femtosecond pulse of XUV/X-ray irradiation, photons excite electrons from core shells and the valence band into the conduction band of the material [22]. As a result, core holes undergo Auger-decays, while the emitted high-energy electrons in the conduction band can excite secondary electrons, forming a cascade of excitation. The interaction between these excited electrons leads to the rapid thermalization of the conduction band electrons at femtosecond timescales [23]. The electrons exchange kinetic energy with the surrounding atoms, which is referred to as the electron-ion (or electron-phonon) coupling.

At the same time, electron excitation also affects the interatomic potential, which, at sufficiently high levels of excitation, may lead to a nonthermal phase transition in covalent materials [24,25]. In this case, bond breaking and atomic rearrangement or disorder occur without significant atomic heating, solely through the changes in the atomic potential energy surface [26]. Over time, the cooling and recombination processes of molecules start to contribute, leading to the relaxation of the atomic system.

The previous work studied the response of polyethylene to FEL irradiation [12]. In the current work, the response of the polypropylene (PP) and the polybutylene (PB-1) to ultrafast XUV and X-ray irradiation is modelled with the hybrid model XTANT-3 [27]. The model includes all the stages of nonequilibrium kinetics, coupling between the electrons and the atoms, nonthermal changes of the atomic potential energy surface due to electronic excitation, and atomic relocation triggered by this change, including nonthermal bond breaking. This enables us to comparatively study the damage processes in alkene series of polymers, identifying their dominant damage mechanisms and thresholds.

## II.    Model

To simulate the damage mechanisms of polyolefins in response to ultrafast X-ray or XUV irradiation in the single shot regime, we use the code XTANT-3 (X-ray-induced Thermal And Nonthermal Transitions) [27]. The XTANT-3 employs the concept of hybrid (multiscale) modelling concurrently executing multiple interlinked models. This allows to efficiently simulate various effects of irradiation [23].

The X-ray photon absorption, nonequilibrium electron kinetics, and secondary cascades are modelled with the transport Monte-Carlo (MC) event-by-event simulation [28]. It includes the kinetics of electrons with energies above a threshold of 10 eV and Auger-decays of core holes. Within the currently achievable fluences at X-ray FELs, single photon absorption is the dominant interaction mechanism [29,30]. Photoabsorption cross sections and Auger decay times are taken from EPICS2023 database [31]. Electron scattering is modelled with the binary-encounter Bethe (BEB) inelastic scattering cross-sections [32]. In this work, MC simulation is averaged over 10,000 iterations to obtain reliable statistics [23,33].

Electrons populating the valence and conduction band with energies below the threshold are assumed to adhere to the Fermi-Dirac distribution. Rate equations are utilized to trace the evolution of the distribution of these electrons on the transient energy levels. The nonadiabatic energy exchange between these electrons and atoms is calculated with the help of the Boltzmann collisional integral [34].

The molecular orbitals representing electron energy levels (band structure) are evaluated with the transferable Tight Binding (TB) method [35]. It involves diagonalization of the transient Hamiltonian, dependent on the position of all the atoms in the simulation box. The gradient of the total energy provided by the Hamiltonian and the transient electronic distribution function describes the interatomic interactions [35]. The wave-function overlap allows us to obtain the matrix element of electron-ion scattering (nonadiabatic coupling), entering the abovementioned Boltzmann collision integral [34]. The transferable TB encompasses parameterizations of hopping integrals and repulsive potentials tailored to pairwise interaction of the atomic species aiming to accurately simulate various material phases [36]. This allows tracing the dynamics and evolution of the system not limited to a single predefined structure. For the alkene polymers, we apply density-function tight binding parameterization matsci-0-3, which uses $sp^3$ basis set for the linear combination of the atomic orbitals of carbon and hydrogen atoms [37].

Atomic motion is traced with the classical Molecular Dynamics (MD), with the forces calculated from the TB and transient electron distribution. The changes in the electronic distribution due to excitation with the XUV/X-ray irradiation directly affect the interatomic potential, and thus may lead to bond breaking and nonthermal melting [12,38]. The nonadiabatic energy transfer, evaluated with the Boltzmann collision integral method, is fed to the atoms *via* the velocity scaling algorithm at each timescale of the simulation [34]. Martyna-Tuckerman 4th order algorithm is used for propagation of the atomic coordinates with the timestep of 0.1 fs [39]. We use 216 atoms in the simulation box with periodic boundary conditions for modelling polypropylene, while 288 atoms are used for polybutylene, sufficient for convergence with respect to the number of atoms [34]. The simulation is run up to 1 ps time.

To identify damage thresholds, we performed two sets of simulations with varying deposited doses from 0.01 eV/atom to 0.1 eV/atom with the step of 0.01 eV/atom, and from 0.1 eV/atom to 1 eV/atom with the step of 0.1 eV/atom, for each material. The dose is defined as the total energy deposited in the simulation box normalized per the total number of atoms in it. The damage here assumes any persistent changes in the structure of the material, global (phase transitions) or local (defect formation, such as dehydrogenation, chain scissions, or cross-linking, in the case of polymers).

We note that the model contains several approximations: electrons are traced with the statistically averaged methods (MC and Boltzmann equation), which provide fractional electronic occupations instead of integer ones. The TB method itself is a crude approximation in comparison with ab initio techniques. The atoms are classical point-like particles in MD, missing quantum effects that may be important for individual molecules. However, as we are attempting to describe the response of highly-excited materials and not fine molecular effects, the methods used here seem to be sufficient, as suggested by previous comparisons with experiments on irradiated materials (see, e.g., Refs. [12,28,40,41]). In particular, XTANT-3 was previously applied for the simulation of polyethylene and poly(methyl methacrylate), showing a reasonable agreement with the experimental data on FEL-induced damage in polymers [12,40]. All the numerical details of the model can be found in Refs. [27,42]. Atomic snapshot visualizations are prepared with the help of OVITO software [43].

## III.    Results

Before productive simulation runs of material response to irradiation, the simulation box with isotactic polymer was relaxed with the steepest descent algorithm. After that, random atomic velocities are initialized corresponding to the Maxwellian distribution, and the Berendsen thermostat is used to thermalize the atomic ensemble at room temperature [44]. Then, we apply the laser pulse of the chosen photon energy of 30 eV, the duration of 10 fs as the full width at half maximum (FWHM) of the Gaussian temporal shape, and various doses to identify where the damage formation starts.

### 1. Polypropylene

The transient electronic cascades are similar to those in polyethylene [12], relaxing within a few femtosecond timescale. Then, the atoms start to respond to the elevated electron temperature, see the example in Figure 1. The electronic temperature peaks during the laser pulse, while the atomic one exhibits an extremely fast increase shortly after. This ultrafast increase of the atomic temperature is not associated with the nonadiabatic electron-ion (electron-phonon) coupling but is a result of the atomic reaction to the nonthermal modification of the interatomic potential. Electronic excitation by the laser pulse changes the interatomic potential to locally repulsive, which triggers atomic acceleration increasing their kinetic energy [45]. The temperatures of the hydrogen and the carbon ensembles coincide during most of the simulation (Figure 1b), apart from the short period during their rise just after the irradiation (~0-50 fs; especially at the highest studied dose). In this short time window before the atomic ensemble equilibration, the hydrogen system is heating up faster than the carbon one, indicating that hydrogen atoms are experiencing larger changes in the interatomic forces, which may lead to hydrogen detachment from the carbon backbone.

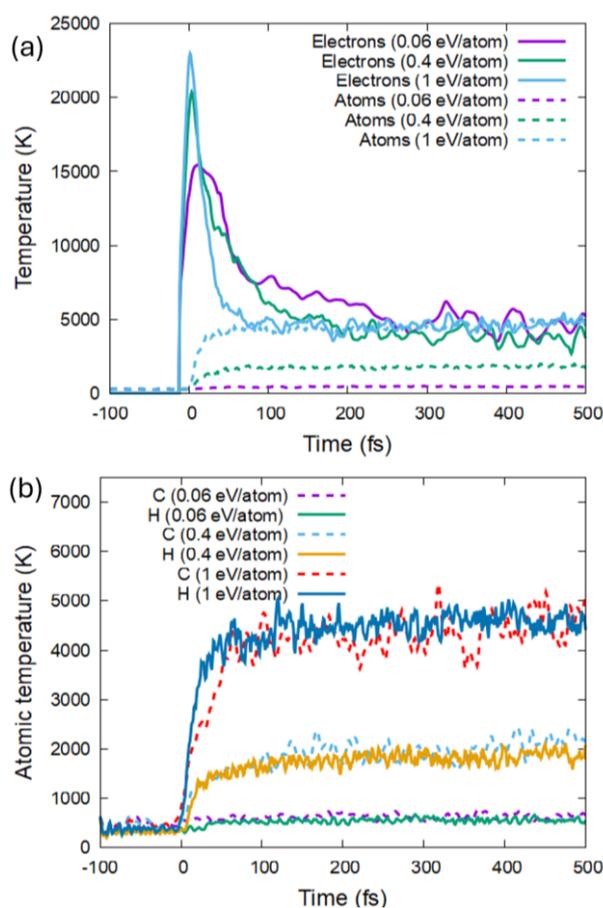

*Figure 1. Evolution of the electronic and the atomic temperatures in PP irradiated with a pulse of 30 eV photon, 10 fs FWHM, and different doses. (a) Electronic and average atomic temperatures. (b) Spicies-specific atomic temperatures.*

Polypropylene (PP) exhibits no sign of damage at doses up to 0.04 eV/atom, whereas persistent defects are produced at the dose of 0.06 eV/atom; The dose of 0.05 eV/atom triggers transient defect formation that may or may not recover in different simulation runs (see examples in Appendix). Thus, we consider that the damage threshold in PP lies at the absorbed dose of ~0.05 ± 0.01 eV/atom, as can be identified by the formation of defect electronic energy levels inside the band gap, see Figure 2 (and Appendix). These defects are associated with dehydrogenation and the formation of hydrogen dimers detached from the backbone of PP, as can be seen in Figure 3 and Appendix.

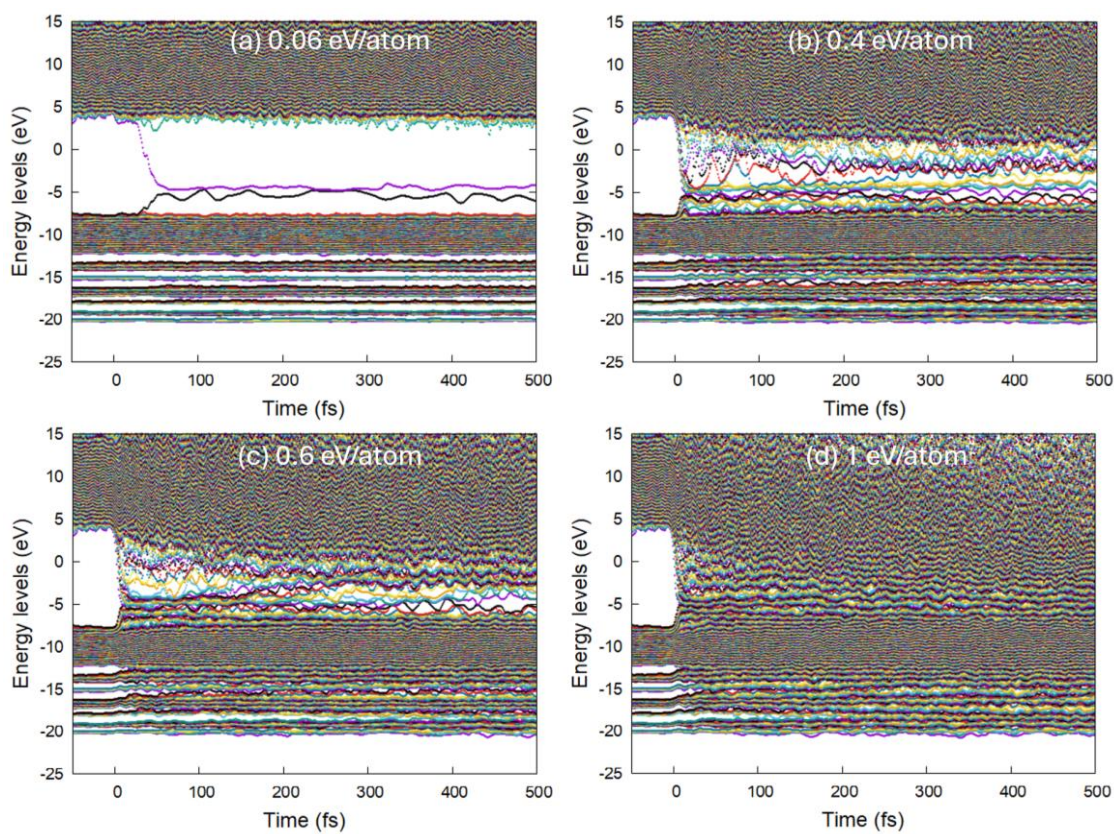

*Figure 2. Electronic energy levels (molecular orbitals) in polypropylene irradiated with 30 eV photons, 10 fs FWHM FEL pulse, (a) the absorbed dose of 0.06 eV/atom; (b) 0.4 eV/atom; (c) 0.6 eV/atom; and (d) 1 eV/atom.*

The detachment of hydrogens is associated with local charge imbalance on the parenting carbon atom (see Appendix). A bond breaking leaves a negatively charged carbon atom behind.

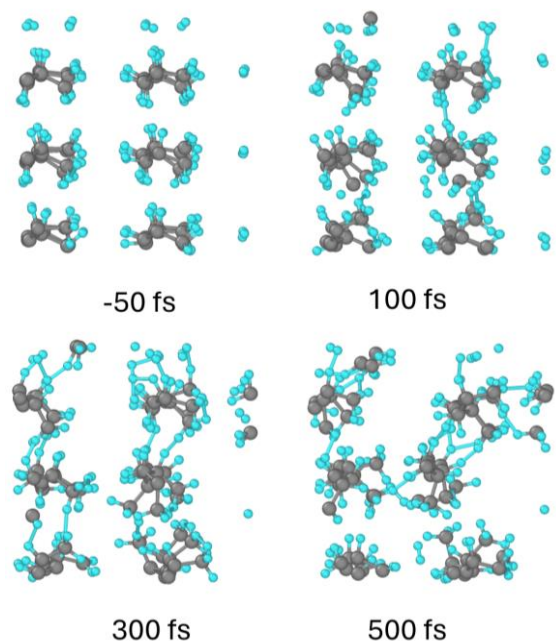

*Figure 3. Atomic snapshots of the PP supercell at different times after irradiation with the dose of 0.06 eV/atom. Grey balls represent carbon atoms, small blue are hydrogens; view along the backbone chains.*

With the increase of the deposited dose, apart from the increase in the dehydrogenation, cross-linking, and scission start to occur (see Figure 4 and Figure 5). In contrast to polyethylene (see Ref.[12]), the next kind of damage is not backbone breaking but $CH_2$ groups detachment from the backbone. An increase in the number of defect energy levels accumulates, and at the dose of ~0.4-0.5 eV/atom, the bandgap completely collapses, merging the valence and the conduction bands of PP (see Figure 2b and c). This corresponds to the transient formation of an electronically conducting, metallic state. This state is characterized by the liquid-like behaviour of the hydrogen subsystem, whereas the carbons are still largely intact in the chains – similar to the previously reported superionic state [12,46]. Formation of a superionic-like state was observed in polyethylene, PMMA (and water), suggesting that it may be a general effect in highly excited polymers [12,40,41].

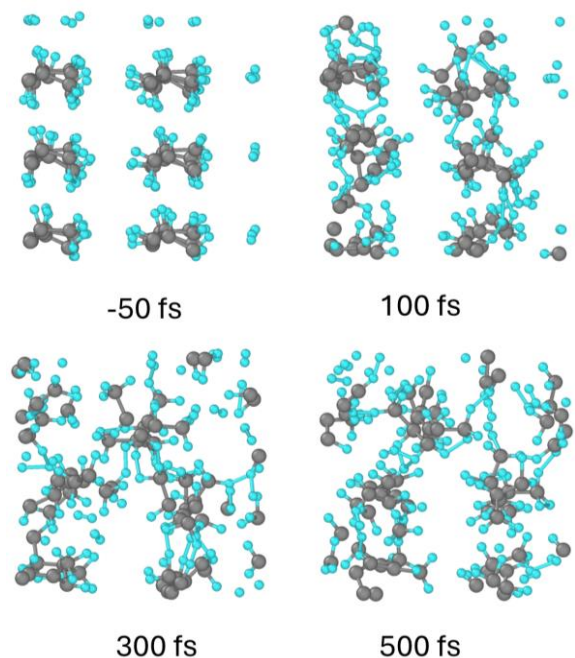

*Figure 4. Atomic snapshots of the PP supercell at different times after irradiation with the dose of 0.4 eV/atom.*

Increasing the absorbed dose to 0.9-1 eV/atom, the backbones of PP are essentially disintegrated, leading to small fragments formed and later complete atomization of material, see Figure 5. It can be expected that in an experiment this will be observed as outgassing, efficient ablation, and plasma formation [40].

All the damage effects – the bandgap collapse, atomic heating, and disordering – occur at the timescale of ~10 fs, significantly faster than the electron-ion coupling leading to the exchange of the kinetic energy and atomic heating. We, thus, conclude that the main driving force behind the damage in ultrafast XUV-irradiated PP is nonthermal bond breaking induced by electronic excitation.

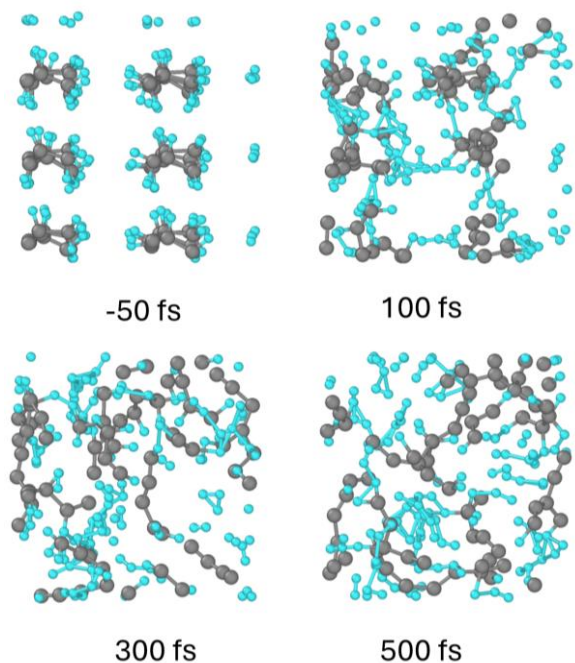

*Figure 5. Atomic snapshots of the PP supercell at different times after irradiation with the dose of 1 eV/atom.*

## 2. Polybutylene

The same series of simulations with XTANT-3 was performed for polybutylene (PB). The results on damage appeared to be similar. The first dehydrogenation takes place at the same dose of ~0.05 eV/atom, as seen by the formation of defect levels in the band structure shown in Figure 6.

With the increase of the dose to ~0.5 eV/atom, $C_2H_4$ groups detach from the carbon backbone, see Figure 7. At this point, metallization of the polybutylene occurs *via* the complete bandgap collapse, the same as for polypropylene presented above. Deposition of ~1 eV/atom triggers the decomposition of the backbone molecules.

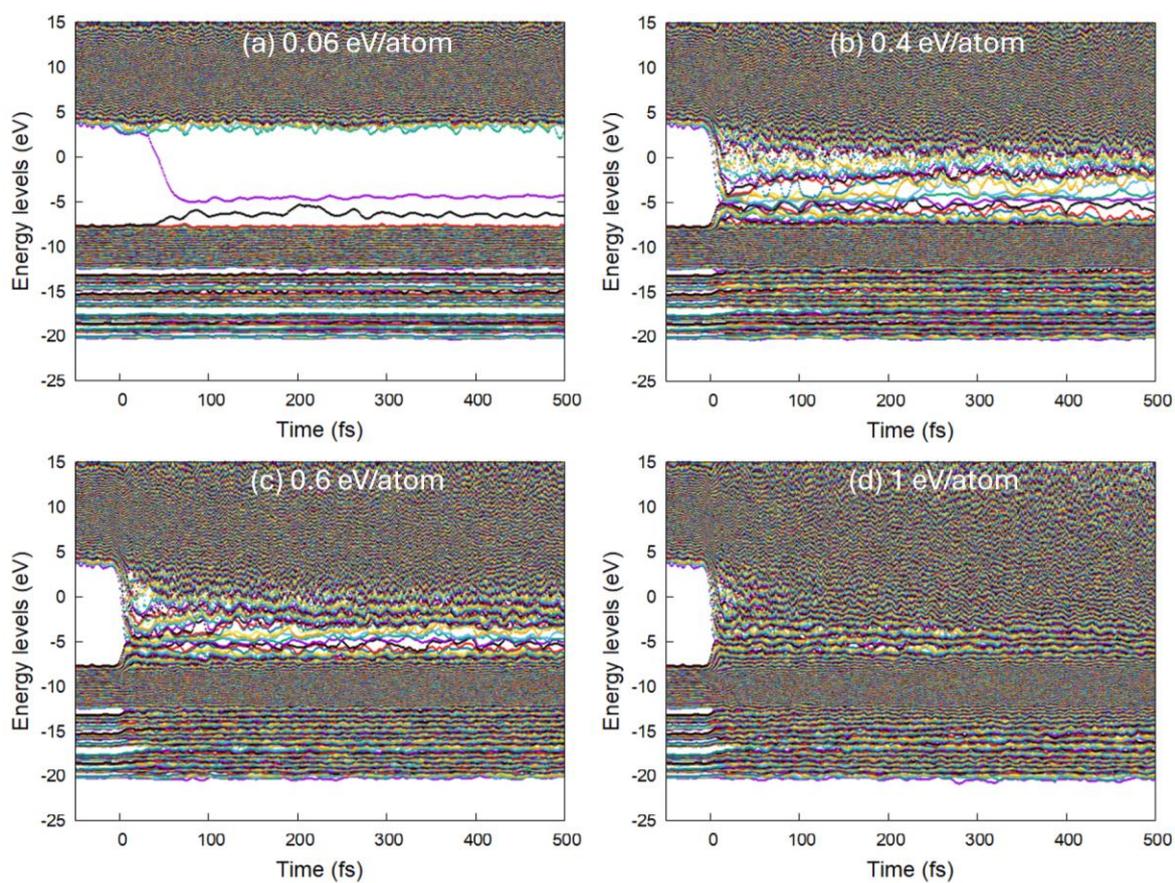

*Figure 6. Electronic energy levels (molecular orbitals) in polybutylene irradiated with 30 eV photons, 10 fs FWHM FEL pulse, (a) the absorbed dose of 0.06 eV/atom; (b) 0.4 eV/atom; (c) 0.6 eV/atom; and (d) 1 eV/atom.*

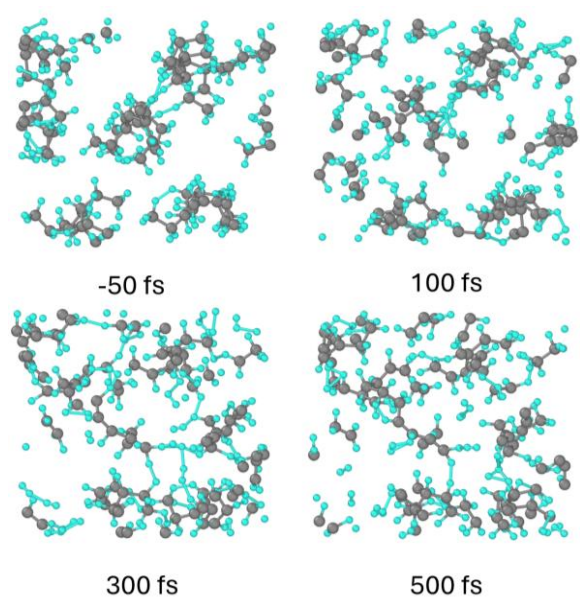

*Figure 7. Atomic snapshots of the PB supercell at different times after irradiation with the dose of 0.4 eV/atom.*

### 3. Damage thresholds in alkene series

Despite drastically different physical, chemical, and mechanical properties among the polyolefins, their response to ultrafast (high dose rate) XUV/X-ray radiation appears to be qualitatively similar (a detailed study of polyethylene was reported in Ref.[12]). Even the quantitative measure, the damage threshold dose, is very close in all the studied cases, related to the same processes taking place: dehydrogenation at the absorbed dose of ~0.05 eV/atom. With the increase of the dose, cross-linking and chain scission occur, resulting in transient metallic state formation at ~0.5 eV/atom, although different groups detach from the backbone depending on the particular material. Breaking of the backbone chains, atomization, and disordering starts at the dose above ~0.9-1 eV/atom.

Generally, the damage threshold depends on the photon energy, temporal and spatial pulse shapes, pulse duration, angle of incidence, and other parameters. Various temporal pulse shapes were studied in Ref. [47], which showed that the shape of the FEL pulse has almost no effect on the processes triggered in the sample, as long as the fluence is below the threshold for multi-phonon effects, and the pulse duration is shorter than the characteristic times of damage and transport effects (ten-femtosecond scale). In the linear absorption regime, all the pulse-shape effects are minor and completely vanish by the end of the pulse.

With an increase in the photon energy, the deeper shells are excited; the photoelectrons acquire higher energies and may be emitted from the surface; the high-energy electron transport, charge imbalance, and local effects may play a role [47]. Those effects are eliminated in our study by design, as we identify an 'intrinsic' damage threshold for the case of homogeneous and uniform excitation of the material. In such a case, the threshold dose may be straightforwardly converted into the incoming threshold fluence under the assumption of the normal incidence of the ultrafast laser pulse as [28]: $F_{th} = D_{th} n_{at} \lambda$ (where $D_{th}$ is the threshold dose, $n_{at}$ is the atomic concentration, and $\lambda$ is the photon attenuation length dependent on the photon energy[31,48]). Figure 8 shows the threshold fluences corresponding to the doses of hydrogen detachment (~0.05 eV/atom), formation of metallic superionic state (~0.5 eV/atom), and complete disorder and atomization (~1 eV/atom). These estimates of the threshold fluences may be used to guide the experiments involving ultrafast XUV/X-ray irradiation of alkene polymers.

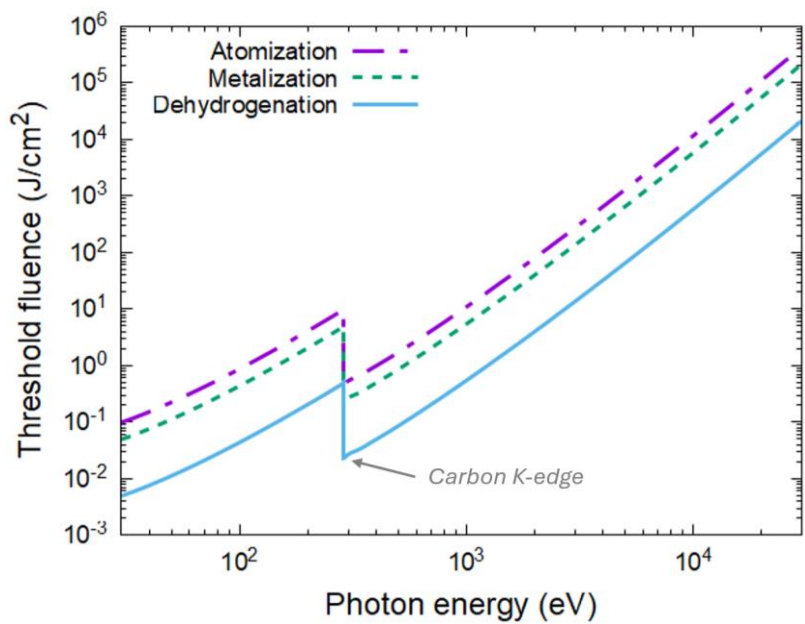

*Figure 8. Damage threshold fluences in alkenes, corresponding to the onset of dehydrogenation (the absorbed dose of 0.05 eV/atom), transient metallization (the dose of ~0.5 eV/atom), and complete disorder or atomization (~1 eV/atom) as the functions of incident photon energy (normal incidence).*

## IV.   Conclusion

Comprehensive modelling of ultrafast XUV/X-ray irradiation of alkene polymers (polypropylene and polybutylene; polyethylene from Ref.[12]) was performed with the help of the XTANT-3 simulation toolkit. The model includes nonequilibrium electron kinetics, nonthermal damage via change of the interatomic potential (bond breaking) induced by electronic excitation, and nonadiabatic electron-ion coupling.

It is shown that the lowest damage threshold dose of ~0.05 eV/atom is associated with the formation of local defects such as dehydrogenation via H‑C bond breaking. With an increasing dose of irradiation to ~0.4 eV/atom, carbon chains begin to actively crosslink while the groups of $CH_2$ and $C_2H_4$ detach from PP and PB, respectively. At such excitation doses, defect levels in the electronic band structure almost completely fill the band gap, leading to transient metallization of the polymers with the formation of a superionic-like state (with liquid hydrogen but structured carbon chains). At higher radiation doses above ~0.9-1 eV/atom, C-C bonds in the backbone chains break and eventually, the entire material disintegrates.

Surprisingly, in all the studied alkene polymers, the ultrafast-X-ray irradiation damage threshold doses are nearly identical. It is associated with the bond energy of hydrogens and carbon chains, which is almost insensitive to a particular structure of the polymer.

## V.    Author contributions (CRediT)

N. Nikishev: conceptualization, investigation, formal analysis, visualization, writing – original draft. N. Medvedev: conceptualization, formal analysis, methodology, software, supervision, visualization, writing – original draft, writing – review & editing.

## VI.    Conflicts of interest

There are no conflicts to declare.

## VII.    Data and code availability

The code XTANT-3 used to produce the data is available from [27].

## VIII.    Acknowledgments

The authors thank L. Juha for fruitful discussions. Computational resources were provided by the e-INFRA CZ project (ID:90254), supported by the Ministry of Education, Youth and Sports of the Czech Republic. NM thanks the financial support from the Czech Ministry of Education, Youth, and Sports (grant nr. LM2023068). NN appreciates the Researchers at Risk Fellowship provided by the Czech Academy of Sciences.

## IX.    Appendix

Atomic snapshots of polypropylene irradiated with the deposited dose of 0.05 eV/atom and 0.06 eV/atom are shown in Figure 9. The atoms in this figure are color-coded according to their Mulliken charge [49]. It can be seen that the carbon with the broken bond acquires a local negative charge. In the example of 0.05 eV/atom, the carbon-carbon bond transiently breaks, forming a scission at the time of ~30-40 fs, which then recovers completely by the time of ~200 fs. This can also be seen in the transient formation and disappearance of the defect electronic energy levels shown in Figure 10 (cf. Figure 2 with the dose 0.06 eV/atom where stable defect levels are formed). In contrast, at the dose of 0.06 eV/atom, hydrogen dimer forms, also leaving a negatively charged carbon atom behind (Figure 9b).

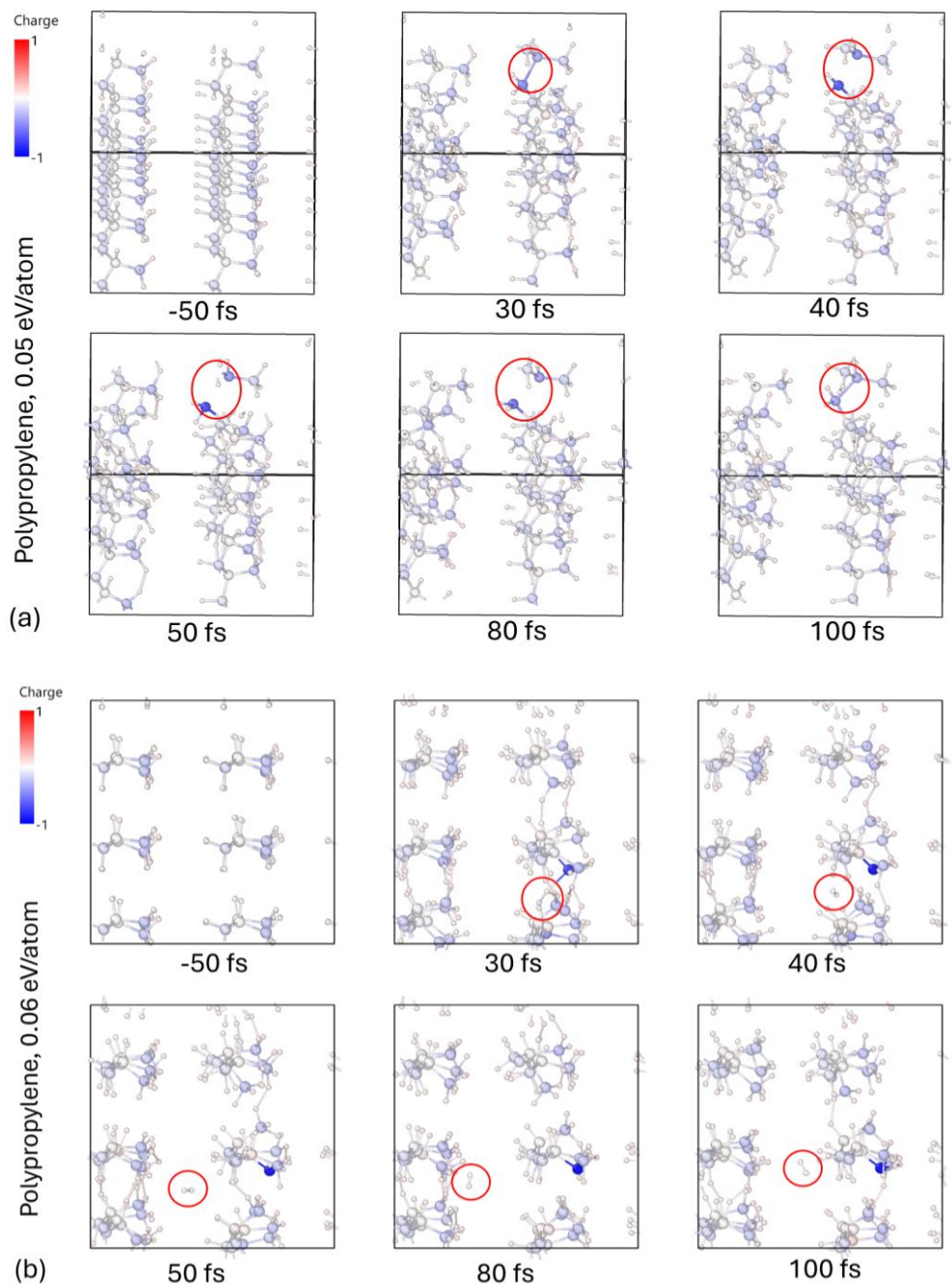

*Figure 9. Atomic snapshots of polypropylene irradiated with 30 eV photons, 10 fs laser pulse with (a) 0.05 eV/atom, and (b) 0.06 eV/atom deposited dose. The atoms are color-coded corresponding to their Mulliken charge. The red circles highlight (a) transient chain scission (bond breaking) and recovery; and (b) dehydrogenation and emission of hydrogen dimer. Large balls are carbon atoms, small are hydrogen.*

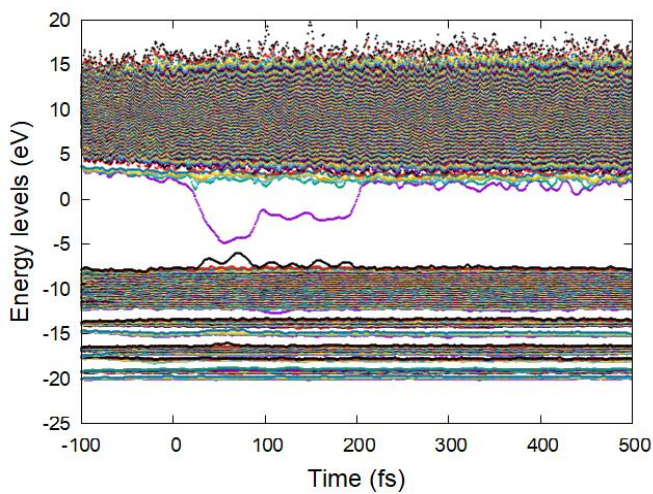

*Figure 10. Electronic energy levels (molecular orbitals) in polypropylene irradiated with 30 eV photons, 10 fs FWHM FEL pulse, the absorbed dose of 0.05 eV/atom.*